\documentclass{aa}  

\usepackage[switch]{lineno}
\usepackage{natbib}
\bibpunct{(}{)}{;}{a}{}{,} 

\usepackage[normalem]{ulem}  
\usepackage{xcolor}          
\usepackage{amssymb}         

\newif\ifhighlight
\highlightfalse  

\ifhighlight
  
  \newcommand{\delRef}[1]{\textcolor{blue}{\textsuperscript{\large\textit{}}}~\textcolor{blue}{\sout{#1}}}

  \newcommand{\delA}[1]{\textcolor{green!50!black}{\textsuperscript{\large\textit{Auth}}}~\textcolor{magenta}{\sout{#1}}}

\else
  
  \newcommand{\delRef}[1]{}

  \newcommand{\delA}[1]{}
  
\fi

\usepackage{graphicx}
\usepackage{txfonts}
\usepackage{placeins}
\usepackage{float}
\begin{document} 


     \title{Quasi-linear approach of bi-Kappa distributed electrons with dynamic $\kappa$ parameter. EMEC instability}
     \titlerunning{Dynamic kappa EMEC instability} 

      \author{Pablo S. Moya \inst{1,2,*}
            \and Roberto E. Navarro\inst{3}
            \and Marian Lazar\inst{2,4}
            \and Peter H. Yoon\inst{5}
            \and Rodrigo A. López\inst{6,7}
            \and Stefaan Poedts\inst{2,8}
          }

    \institute{Departamento de F\'isica, Facultad de Ciencias, Universidad de Chile, Las Palmeras 3425, \~{N}u\~{n}oa, Santiago 7800003, Chile\\
               \email{pablo.moya@uchile.cl}
               \and
               Center for mathematical Plasma Astrophysics, KU Leuven, Celestijnenlaan 200B, B-3001 Leuven, Belgium
               \and
               Departamento de F\'isica, Facultad de Ciencias F\'isicas y Matem\'aticas, Universidad de Concepci\'on, Concepci\'on, Chile
               \and
               Institut f\"{u}r Theoretische Physik, Lehrstuhl IV: Weltraum- und Astrophysik, Ruhr-Universität Bochum, D-44780 Bochum, Germany
               \and
               Institute for Physical Science and Technology, University of Maryland, College Park, MD 20742-2431, USA
               \and 
               Research Center in the intersection of Plasma Physics, Matter, and Complexity ($P^2 mc$), Comisi\'on Chilena de Energ\'{\i}a Nuclear, Casilla 188-D, Santiago, Chile
               \and 
               Departamento de Ciencias F\'{\i}sicas, Facultad de Ciencias Exactas, Universidad Andres Bello, Sazi\'e 2212, Santiago 8370136, Chile
               \and
               Institute of Physics, University of Maria Curie-Sk{\l}odowska, Pl.\ M.\ Curie-Sk{\l}odowska 5, 20-031 Lublin, Poland.
               }

\date{Received ; accepted }

  \abstract
   {In recent years, significant progress has been made in the velocity-moment-based quasi-linear (QL) theory of waves and instabilities in plasmas with nonequilibrium velocity distributions (VDs) of the Kappa (or $\kappa$) type. However, the temporal variation of the parameter $\kappa$, which quantifies the presence of suprathermal particles, is not fully captured by such a QL analysis, and typically $\kappa$ remains constant during plasma dynamics.}
   {We propose a new QL modeling that goes beyond the limits of a previous approach, realistically assuming that the quasithermal core cannot evolve independently of energetic suprathermals.}
   {The case study is done on the electron-cyclotron (EMEC) instability generated by anisotropic bi-Kappa electrons with $A=T_\perp/T_\parallel > 1$ ($\parallel, \perp$ denoting directions with respect to the background magnetic field). The parameter $\kappa$ self-consistently varies through the QL equation of kurtosis (fourth-order moment) coupled with temporal variations of the temperature components, relaxing the constraint on the independence of the low-energy (core) electrons and suprathermal high-energy tails of VDs.}
   {The results refine and extend previous approaches. A clear distinction is made between regimes that lead to a decrease or an increase in the $\kappa$ parameter with saturation of the instability. What predominates is a decrease in $\kappa$, i.e., an excess of suprathermalization, which energizes suprathermal electrons due to self-generated wave fluctuations. Additionally, we found that VDs can evolve toward a quasi-Maxwellian shape (as $\kappa$ increases) primarily in regimes with low beta and initial kappa values greater than five.}
  {Instability-driven relaxation only partially resolves temperature anisotropy in bi-Kappa electron VDs, as wave fluctuations generally act to further energize suprathermal electrons. The present results show a preliminary agreement with in situ observations in the solar wind, suggesting that the new QL model could provide a sufficiently explanatory theoretical basis for the kinetic instabilities in natural plasmas with Kappa-like distributions.}

   \keywords{Vlasov quasi-linear theory, Whistler electron-cyclotron instability, Kappa distributions}

   \maketitle

   \FloatBarrier

\section{Introduction} 
\label{sec:intro}

In the solar wind, the velocity distributions (VDs) deviate from the idealized Maxwellian profile specific to thermal equilibrium, and exhibit several nonthermal features, such as temperature anisotropy relative to the magnetic field direction, and suprathermal populations as high-energy tails decreasing as Kappa (or $\kappa$) power laws of the velocity. Introduced as an empirical model to reproduce the electron VDs observed in the terrestrial magnetosphere \citep{Olbert-1968, Vasyliunas-1968}, and later the distributions of all plasma species in the solar wind \citep{Christon-etal-1989, Christon-etal-1991, Collier-etal-1996, Maksimovic-etal-1997}, Kappa models have become a revolutionary kinetic tool in the last decades, enabling nonequilibrium plasma states to be described and understood \citep{Stverak-etal-2008, Vinas-etal-2015, Lazar-etal-2015, Dzifcakova-etal-2018, Yoon-2021, Shaaban-etal-2021, Lopez_Shaaban_Lazar_2021, Moya-etal-2021}. 

Since space plasmas are weakly collisional, self-generated wave instabilities are expected to control the kinetic anisotropies of the VDs. Observations seem to confirm this hypothesis, such as in the case of temperature anisotropy $A= T_\perp / T_\parallel \ne 0$ (relative to the direction of the magnetic field), and especially for quasi-thermal core populations at low energies \citep{Stverak-etal-2008, Bale-etal-2009, Shaaban-etal-2025}. The observed distributions are sufficiently gyrotropic, and one can invoke bi-Maxwellian models to describe the core populations with temperature anisotropy, and bi-Kappa models for anisotropic distributions with suprathermal tails \citep{Summers-Thorne-1991, Stverak-etal-2008, Lazar-etal-2017}.
For bi-Kappa populations, a refined analysis is still necessary \citep{Moya-etal-2021, Shaaban-etal-2025}, to account for not only the relaxation of the anisotropy, but also the modifications suffered by the suprathermal tails via the parameter $\kappa$, as revealed by numerical simulations \citep{Lazar-etal-2018, Shaaban-etal-2024}. This paper proposes a new quasi-linear (QL) approach that generalizes the zero-order QL approach in \cite{Moya-etal-2021}. The QL theory has the ability to explain the results observed in numerical simulations, in particular the mechanisms by which the growth and stabilization of instabilities are achieved, under the action of the wave fluctuations thus generated. We do this for the concrete case of whistler instability, also known as electromagnetic electron-cyclotron (EMEC) mode instability, triggered by anisotropic electrons with $A > 1$. However, the new approach proposed here can easily be extended to instabilities of other EM modes, such as the firehose instability (driven by $A<1$), as well as to instabilities driven by anisotropic protons and ions.

Our paper is structured as follows. In Sect.~2 we construct the new first-order QL formalism of the whistler instability, which goes beyond the limits of a zero-order approach \citep{Moya-etal-2021}, realistically assuming that the quasithermal electron core is still dominant, but cannot evolve independently of the suprathermal component. The new QL model is based on the kurtosis components, introduced as fourth-order moments of the Kappa distribution, which allow us to treat the temporal variation of the parameter $\kappa (t)$ in a self-consistent manner. The parametric analysis in Sect.~3 is performed on the results obtained for various initial conditions, by varying the key parameters, the temperature anisotropy ($A$), the plasma beta parameter (parallel component, $\beta_\parallel$), and the parameter $\kappa$. The last section, Sect.~4, draws the main conclusions and analyzes a series of perspectives opened by these results.

\section{Quasi-linear model}
\label{sec:theory}
%
For our model, we consider a quasi-neutral magnetized plasma composed of electrons and cold ions, in which electrons follow a bi-Kappa VD given by 

\begin{equation}
f_0(v_\perp,v_\parallel) =\dfrac{n_0}{\pi^{3/2}\theta_\perp^2\theta_\parallel}
\dfrac{\Gamma(\kappa)}{\kappa^{1/2}\Gamma(\kappa-1/2)} \Biggl(1+\dfrac{v_\perp^2}{\kappa\theta_\perp^2}
+\dfrac{v_\parallel^2}{\kappa\theta_\parallel^2}\Biggr)^{-\kappa-1} 
\label{eq:VDF}    
,\end{equation}
where $n_0$ is the total number density and $\kappa$ is the power-law index representing the presence of suprathermal particles. The $\kappa$ index increases for decreasing high-energy tails, and $\kappa \to \infty$ corresponds to a Maxwellian VD. In Eq.~\eqref{eq:VDF} the $\theta_\perp$ and $\theta_\parallel$ parameters define the parallel, $T^{(\kappa)}_\parallel$, and perpendicular, $T^{(\kappa)}_\perp$, kinetic temperatures by the second-order
moment of the distribution function. Namely, 

\begin{eqnarray}
\label{eq:Tpal}
M_{2\parallel} = k_B T^{(\kappa)}_\parallel  &=&  \int \, m_e\, v^2_\parallel\, f_0(v_\perp,v_\parallel)\,d\mathbf{v} = \dfrac{m_e\theta_\parallel^2}{2} \dfrac{\kappa}{\kappa-3/2}\\
\label{eq:Tper}
M_{2\perp} = k_B T^{(\kappa)}_\perp  &=&  \int \dfrac{1}{2}\, m_e\, v^2_\perp\, f_0(v_\perp,v_\parallel)\,d\mathbf{v}\,=\dfrac{m_e\theta_\perp^2}{2} \dfrac{\kappa}{\kappa-3/2}\,,
\end{eqnarray}
where $k_B$ is the Boltzmann constant and $m_e$ the electron mass. We can also interpret the $\theta_j$ parameters as the thermal speeds of a bi-Maxwellian distribution with temperatures given by
\begin{equation}
\theta_\parallel^2 = \dfrac{2 k_B T_\parallel}{m_e} ,\qquad  \theta_\perp^2 = \dfrac{2 k_B T_\perp}{m_e}\,.
\label{eq:theta}
\end{equation}
Thus, combining Eqs.~\eqref{eq:Tpal}, \eqref{eq:Tper}, and~\eqref{eq:theta}, the relation between the kinetic and Maxwellian temperatures is the following:
\begin{equation}
T^{(\kappa)}_\parallel=F(\kappa)\,T_\parallel,\qquad  
T^{(\kappa)}_\perp=F(\kappa)\,T_\perp\,
\label{eq:TandTheta}
,\end{equation}
where
\begin{equation}
F(\kappa) = \dfrac{\kappa}{\kappa-3/2}    
\label{eq:FK}
.\end{equation}

\subsection{Whistler-cyclotron dispersion relation}

Under this model, the Vlasov dispersion relation for right-hand polarized waves propagating in the direction of the background magnetic field, $\mathbf{B_0}$, is given by~\citep{Vinas-etal-2015, Lazar-etal-2017b, Moya-etal-2021}
\begin{equation}
0=\dfrac{c^2k_\parallel^2}{\omega_{pe}^2}-\left(A-1\right)-\dfrac{A \omega
-\left(A-1\right)|\Omega_e|}{k_\parallel\theta_\parallel}\,Z_\kappa
\left(\dfrac{\omega-|\Omega_e|}{k_\parallel\theta_\parallel}\right)\,, \label{eq:disp}
\end{equation}
where $\omega$ and $k_\parallel$ are the wave frequency and wavenumber, respectively, and $\omega_{pe} = \sqrt{4\pi n_0 e^2/m_e}$ and $\Omega_{e} = e B_0/m_e c$ are the electron plasma frequency and gyrofrequency. Also, in Eq.~\eqref{eq:disp} $e$ is the elementary charge, $c$ is the speed of light, and
\begin{equation}
Z_\kappa(\xi)=\dfrac{1}{\pi^{1/2}\kappa^{1/2}} \dfrac{\Gamma(\kappa)}{\Gamma(\kappa-1/2)}\int_{-\infty}^\infty
dx\,\dfrac{(1+x^2/\kappa)^{-\kappa}}{x-\xi},\qquad {\rm Im}(\xi)>0 \label{eq:Z}
\end{equation}
is the modified dispersion function~\citep{Hellberg-Mace-2002, Lazar-etal-2008}. It is important to mention that in the Maxwellian limit, $\kappa \to \infty$, $Z_k$ becomes the well-known plasma dispersion function, $Z$, defined by~\citet{Fried-Conte-1961}.

For temperature anisotropy 
\begin{equation}
A = \dfrac{T^{(\kappa)}_\perp}{T^{(\kappa)}_\parallel} = \dfrac{\theta_\perp^2}{\theta_\parallel^2} = \dfrac{T_\perp}{T_\parallel} > 1\,, \label{eq:aniso}
\end{equation}
the dispersion relation Eq.~\eqref{eq:disp} is unstable to whistler EMEC waves, with a maximum growth rate that increases with increasing temperature anisotropy, $A>1$, and increasing parallel plasma beta, $\beta_\parallel = 8\pi n_0 k_B T_\parallel/B^2_0$~\citep[see e.g.][and references therein]{gary1993}. This behavior is similar to that of a collisionless plasma composed of anisotropic electrons, and, for Kappa-distributed electrons, several studies have already been conducted~\citep{Vinas-etal-2015, Lazar-etal-2017, Lazar-etal-2018}. 
In the applications, we consider combinations of temperature anisotropy and plasma beta, and for each combination various values of $\kappa$. For comparison between Kappa distributions and the Maxwellian limit ($\kappa \to \infty$), we consider the kinetic beta parameter
\begin{equation}
\beta^{(\kappa)}_{\parallel} = \dfrac{8 \pi n_0 k_B T^{(\kappa)}_\parallel}{B^2_0} = F(\kappa)\,\dfrac{8 \pi n_0 k_B T_\parallel}{B^2_0}= F(\kappa)\,\beta_\parallel,\label{eq:beta}
\end{equation}
which reduces to the lower Maxwellian limit $\beta_{\parallel}^{(\kappa)}  \to \beta_{\parallel} < \beta_{\parallel}^{(\kappa)}$, when $\kappa \to \infty$.

\subsection{QL evolution of power-index $\kappa$}

One of the most traditional ways to include nonlinear effects of kinetic instabilities is the use of QL theory~\citep[see e.g.][and references therein]{Yoon-2017}, which provides expressions for the time evolution of the plasma VD and electromagnetic fluctuations due to nonlinear interactions
\begin{eqnarray}
    \dfrac{\partial f_0}{\partial t} &= \dfrac{i\,e^2}{4\,m^2_e\,c^2}\int^{\infty}_{-\infty} \dfrac{dk_\parallel}{k^2_\parallel}\left[(\omega^* - k_\parallel v_\parallel)\dfrac{\partial}{\partial v_\perp}+k_\parallel v_\perp \dfrac{\partial}{\partial v_\parallel}\right] \nonumber\\
    & \times\dfrac{v^2_\perp \delta B^2(k_\parallel)}{\omega-k_\parallel v_\parallel+\Omega_e} \left[(\omega - k_\parallel v_\parallel)\dfrac{\partial f_0}{\partial v_\perp}+k_\parallel v_\perp \dfrac{\partial f_0}{\partial v_\parallel}\right]\label{eq:dfdt}
,\end{eqnarray}
where $\omega^*$ represents the complex conjugate of the complex solutions of the dispersion relation Eq.~\eqref{eq:disp}, $\omega = \omega(k_\parallel) +i\gamma(k_\parallel)$, and $\delta B^2(k_\parallel)$ is the wave energy density whose temporal evolution is given by the wave kinetic equation 
\begin{equation}
\dfrac{\partial\delta B^2(k_\parallel)}{\partial t} =2\gamma(k_\parallel)\,\delta B^2(k_\parallel)\,. \label{eq:dBdt}
\end{equation}

Then, taking the temporal derivative on both sides of Eqs.~\eqref{eq:Tpal} and \eqref{eq:Tper}, and using Eq.~\eqref{eq:dfdt}, from QL theory we obtain
\begin{eqnarray}
k_B \dfrac{dT^{(\kappa)}_\parallel}{dt} &=&  \int \dfrac{1}{2}\, m_e\, v^2_\parallel\, \dfrac{\partial f_0}{\partial t}\,d\mathbf{v} \nonumber \\
&=& \dfrac{e^2}{m_e} \int_{-\infty}^\infty\dfrac{dk_\parallel}{c^2k_\parallel^2}
\,\gamma(k_\parallel)\left(\dfrac{c^2k_\parallel^2} {\omega_{pe}^2}+1\right)\delta B^2(k_\parallel)\,, \label{eq:dTparadt}\\
k_B \dfrac{dT^{(\kappa)}_\perp}{dt} &=&  \int  m_e\, v^2_\perp\, \dfrac{\partial f_0}{\partial t}\,d\mathbf{v}   \nonumber \\ 
&=& -\dfrac{e^2}{m_e}\int_{-\infty}^\infty \dfrac{dk_\parallel}{c^2k_\parallel^2}\,\gamma(k_\parallel)
\left(\dfrac{c^2k_\parallel^2}{\omega_{pe}^2} +\dfrac{1}{2}\right)\delta B^2(k_\parallel). \label{eq:dTperpdt}
\end{eqnarray}

Equations~\eqref{eq:dTperpdt} and \eqref{eq:dTparadt}, together with the wave kinetic equation, Eq.~\eqref{eq:dBdt}, and the dispersion relation, Eq.~\eqref{eq:disp}, provide a closed system of differential equations that describe the time evolution and saturation of the EMEC instability in terms of the coupled variation of the plasma temperatures and the wave energy. From the expressions given in~\eqref{eq:TandTheta}, their variation in time (i.e., $dT^{(\kappa)}_{\parallel, \perp}/dt \neq 0$) implies variations in time not only for $T_{\parallel, \perp}$ (as assumed in a simplified model) but also for the exponent $\kappa$. In previous studies, it has been customary to consider $\kappa = \kappa_0$ =  constant, as is assumed in several QL approaches with $\kappa$ distributions~\citep[see e.g][]{Lazar-etal-2017, Lazar-etal-2018}. Thus, in such cases, Eqs.~\eqref{eq:disp},~\eqref{eq:dBdt}, and~\eqref{eq:dTperpdt}-\eqref{eq:dTparadt} provide a complete QL description of a temperature anisotropy instability and its relaxation. 

On the other hand, if $\kappa$ changes in time, i.e., $\kappa = \kappa(t)$, as is suggested by the Vlasov and PIC simulations in~\citet{Lazar-etal-2017b} and~\citet{Lazar-etal-2018}, respectively, we require an additional independent equation -- different from but complementary to Eqs.~\eqref{eq:dTperpdt} and \eqref{eq:dTparadt} -- to account for the time evolution of $\kappa$. Since $f_0$ is symmetric in the magnetic field direction, all odd moments are identically zero. Therefore, the additional equation can be obtained by considering the fourth-order moment (the smallest non-null moment besides the kinetic temperature and number density) of the Kappa VD, which is related to the kurtosis of the distribution. The kurtosis, $K$, is a well-known measure of the relative weight of the high-energy tails of the VD with respect to the core; namely,
\begin{align}
K =  \int v^4 \,  f_0(v_\perp,v_\parallel) \,d\mathbf{v} = &\left(\dfrac{\kappa}{\kappa-3/2}\right)\left(\dfrac{\kappa}{\kappa-5/2}\right) \nonumber \\
& \times \left(2\theta^4_\perp +\theta^2_\perp\theta^2_\parallel+\dfrac{3}{4}\theta^4_\parallel\right). \label{eq:kurt}
\end{align}
Note that, for the convergence of the fourth-order moment, $\kappa > 5/2 $ is required, a condition that we also take into account when obtaining and analyzing the results. Note that the adoption of the regularized Kappa distribution eliminates the need to restrict the Kappa index~\citep{Scherer2017, Thanh-2019b}. However, for the sake of simplicity, the present study does not take such an additional feature into account.

In terms of temperatures, using Eq.~\eqref{eq:theta} and \eqref{eq:TandTheta}, the kurtosis is given by
\begin{align}
K(t) = \int v^4 \,  & f_0(v_\perp,v_\parallel) \,d\mathbf{v} = \dfrac{4k^2_B}{m_e^2} \dfrac{\kappa(t)}{\kappa(t)-3/2} \dfrac{\kappa(t)}{\kappa(t)-5/2} \nonumber \\
& \times \left[2\, (T_\perp(t))^2 +T_\perp(t)\,T_\parallel(t)+\dfrac{3}{4} (T_\parallel(t))^2 \right]. \label{eq:kurtT}
\end{align}
Therefore, in the case of Kappa VDs, the kurtosis is a measure or indicator of the shape of the distribution and is highly controlled by the $\kappa$ parameter. Thus, considering the time derivative of kurtosis $K$ provides a way to obtain a dynamic equation for the time evolution of $\kappa$. Indeed, taking the temporal derivative on both sides of Eqs.~\eqref{eq:Tpal},~
\eqref{eq:Tper}, and ~\eqref{eq:kurtT}, and using Eq.~\eqref{eq:dfdt}, from QL theory we obtain the following system of coupled differential equations:
\begin{eqnarray}
\label{eq:ql2pal}
\left(\dfrac{m_e}{k_B}\right)\,\dfrac{d M_{2\parallel}}{dt} &=& F(\kappa)\,\dfrac{d T_\parallel}{dt} + T_\parallel \dfrac{dF}{d\kappa}\,\dfrac{d\kappa}{dt}\\
\label{eq:ql2per}
\left(\dfrac{m_e}{k_B}\right)\,\dfrac{d M_{2\perp}}{dt} &=& F(\kappa)\,\dfrac{d T_\perp}{dt} + T_\perp \dfrac{dF}{d\kappa}\,\dfrac{d\kappa}{dt}\\
\left(\dfrac{m_e}{2\,k_B}\right)^2 \dfrac{dK}{dt} =&&
G(\kappa)\left(4T_\perp+T_\parallel\right)\,\dfrac{dT_\perp}{dt} \nonumber \\
&& + G(\kappa)\left(T_\perp+\dfrac{3}{2}\,T_\parallel\right)\,\dfrac{dT_\parallel}{dt} \nonumber \\ 
&&+\left(2\,T_\perp^2+T_\perp\,T_\parallel+\dfrac{3}{4}\,T_\parallel^2\right)\dfrac{dG}{d\kappa}\,\dfrac{d\kappa}{dt}\,, \label{eq:ql4}
\end{eqnarray}
where
\begin{eqnarray}
\label{eq:dMpaldt}
\dfrac{dM_{2\parallel}}{dt} &=& \phantom{-}\dfrac{\omega^2_{pe}}{m_e n_0}\int_{-\infty}^\infty \dfrac{dk_\parallel}{c^2k_\parallel^2}\,\gamma(k_\parallel)
\left(\dfrac{c^2k_\parallel^2}{\omega_{pe}^2} +1\right)\delta B^2(k_\parallel)\\
\label{eq:dMperdt}
\dfrac{dM_{2\perp}}{dt} &=& -\dfrac{\omega^2_{pe}}{m_e n_0}\int_{-\infty}^\infty \dfrac{dk_\parallel}{c^2k_\parallel^2}\,\gamma(k_\parallel)
\left(\dfrac{c^2k_\parallel^2}{\omega_{pe}^2} +\dfrac{1}{2}\right)\delta B^2(k_\parallel)\\
\label{eq:dKdt}
\dfrac{dK}{dt} &=& \phantom{-} \dfrac{4k_B T_\parallel\,\omega^2_{pe}}{m^2_e n_0}\,
\dfrac{\kappa^{1/2}(\kappa-1)^{1/2}}{\kappa-3/2}\int_{-\infty}^\infty \dfrac{dk_\parallel}{c^2k_\parallel^2}\, {\rm{Im}}\left\{\left[\dfrac{|\omega|^2}{k_\parallel\,\theta_\parallel} \right.\right. \nonumber \\ && 
\left. \left. +\left(\dfrac{\kappa}{\kappa-1}\right)^{1/2}\left(\dfrac{T_\perp}{T_\parallel}-1\right)\,\omega^*\xi_\kappa\right]Z_{\kappa-1}(\xi_\kappa)\right\}\delta B^2(k_\parallel)\nonumber\\
&&+\dfrac{4k_B T_\parallel\,\omega^2_{pe}}{m^2_e n_0}\int_{-\infty}^\infty \dfrac{dk_\parallel}{c^2k_\parallel^2}\,\gamma(k_\parallel)\left[\left(\dfrac{\omega^2(k_\parallel)+\gamma^2(k_\parallel)-\Omega^2_e}{k^2_\parallel\,\theta^2_\parallel}\right) \right. \nonumber \\
&& \left.\times \dfrac{c^2k_\parallel^2}{\omega_{pe}^2} - \dfrac{1}{2}\left(\dfrac{\kappa}{\kappa-3/2}\right)\left(\dfrac{T_\perp}{T_\parallel}-1\right)\left(1+4\,\dfrac{T_\perp}{T_\parallel}\right)\right.\nonumber\\
&& + \left. \dfrac{\omega^2 (k_\parallel) + \gamma^2(k_\parallel)}{k^2_\parallel\,\theta^2_\parallel}\right] \delta B^2(k_\parallel). \label{k4}
\end{eqnarray}
Here, 
\begin{align}
\xi_\kappa = \left(\dfrac{\kappa}{\kappa-1}\right)^{1/2}\left(\dfrac{\omega-|\Omega_e|}{k_\parallel\theta_\parallel}\right),
\end{align}
$F(\kappa)$ is given by Eq.~\eqref{eq:FK}, and
\begin{equation} \label{eq:fgkappa}
G(\kappa) = \left(\dfrac{\kappa}{\kappa-3/2}\right)\left(\dfrac{\kappa}{\kappa-5/2}\right)\,.
\end{equation}

In summary, Eq.~\eqref{eq:ql4} provides a new independent equation to close the system formed by Eqs.~\eqref{eq:disp},~\eqref{eq:dBdt},~\eqref{eq:ql2pal}, and~\eqref{eq:ql2per}, allowing the self-consistent variation of the $\kappa$ parameter due to the relaxation of electron VD under the action of EMEC instability. It is important to mention that energy conservation is guaranteed within our QL approach. Indeed, although Eqs.~\eqref{eq:ql2pal}-\eqref{k4} go beyond the typical QL picture, the number of particles, momentum, and energy conservation are always satisfied in QL theory, regardless of the shape of the VD. A proof of these facts can be found in plasma physics textbooks \citep[see e.g. Chapter 10 in][]{krall1973}.

\section{Results}
\label{sec:results}
\begin{table}[t!]
\centering
\caption{Initial electron plasma parameters.}
\begin{tabular}{ccc}
\hline
             $\beta_\parallel (t=0)$ & $\kappa (t=0)$ & $A(t=0)$\\
        \hline
          0.10 & 2.6& 2.00 \\
          0.13 & 2.8& 2.49 \\
          0.16 & 3.0& 3.10 \\
          0.20 & 4.0& 3.87 \\
          0.25 & 5.0& 4.82 \\
          0.50 & 6.0& 6.00 \\
          1.00 & 7.0& 7.47 \\
          2.00 & 8.0&  \\
          4.00 &  &  \\
          8.00 &  &  \\
          \hline
    \end{tabular}
    \label{tab:parameters}
\end{table}

To solve the QL equations, we employed a discrete grid within the wavenumber space, normalized to the electron inertial length $c/\omega_{pe}$, featuring 256 points ranging from $0.001 <|ck_\parallel|/\omega_{pe} < 3$. 
All cases were simulated up to $|\Omega_e| t_{\rm end} = 655$ (or $2^{15}$ time steps) using a time step of $\Delta t = 0.02/|\Omega_e|$. The magnetic field spectrum was specified as $\delta B_k^2/B^2_0 = 10^{-5}$ at $t = 0$. 
Following this methodology, and knowing the magnetic field spectrum and the value of all parameters ($A$, $\beta_\parallel$, and $\kappa$) at any given time, $t$, we solved the dispersion relation to determined the complex frequency, $\omega + i\gamma$, of whistler-cyclotron waves as a function of $k_\parallel$ at that specific moment. Subsequently, we evaluated the time derivative of each parameter. The entire system was then advanced to the next time, $t + dt$, using a second-order Runge-Kutta method. Furthermore, in all cases, we considered $\omega_{pe}/|\Omega_e|=20$, $m_p/m_e=1836$, and cold ions, but in each case, different initial conditions ($t_0=0$) were set in terms of the key electron parameters. This computational strategy was applied to solve the QL system using 560 combinations of $\kappa$, temperature anisotropy, and parallel beta as initial conditions, as is shown in Table~\ref{tab:parameters}.

\begin{figure*}[t!] \centering
\includegraphics[width=0.9\textwidth]{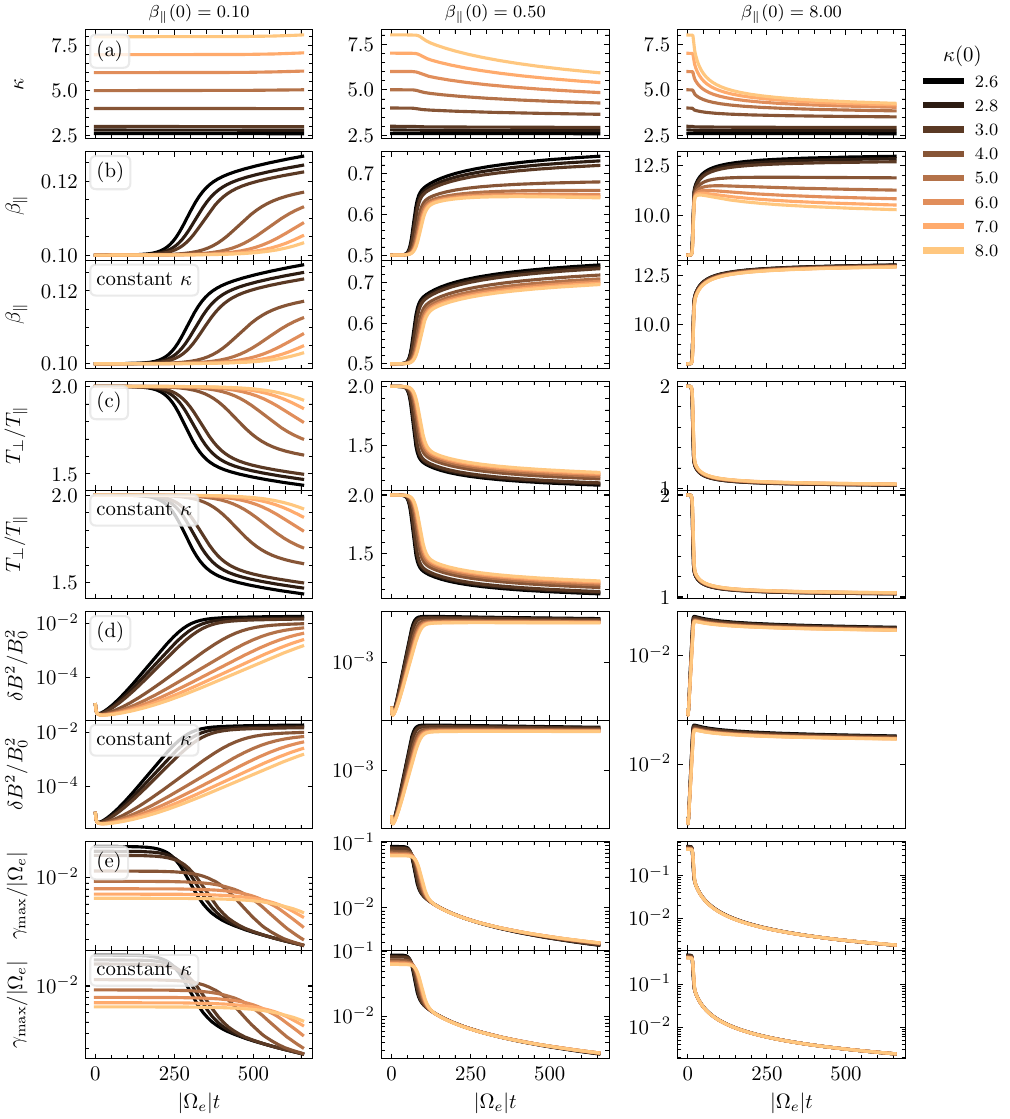}
\caption{\label{fig:1} QL runs with the same initial anisotropy, $A(0)=2$, three values of initial $\beta_\parallel (0)$ = 0.1 (left), 0.5 (middle), and 8.0 (right), and for eight different (initial) values of $\kappa (0)$ shown in various shades of brown. The first line (top) indicates the variation in $\kappa$, and the following ones indicate the variations in key parameters $\beta_\parallel (t)$ and $A (t)=T_\perp (t)/T_\parallel (t)$, and variations in wave properties $\delta B^2 (t)/B_0^2$ and $\gamma_{\rm max} (t)$, for both cases in which $\kappa$ varies in time and ones in which $\kappa$ is constant.}
\end{figure*}

\begin{figure*}[t!] \centering
\includegraphics[width=0.9\textwidth]{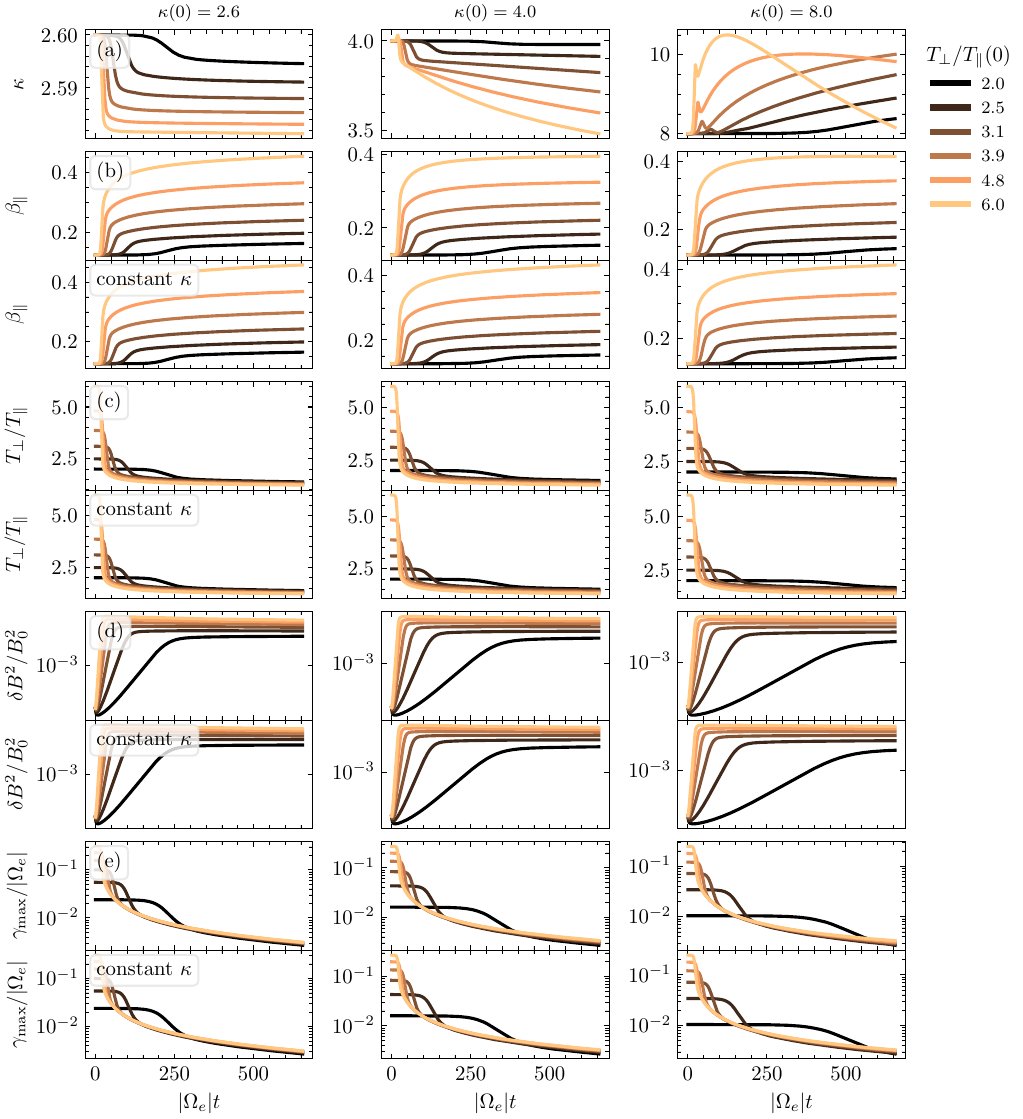}
\caption{\label{fig:2} QL runs with the same initial $\beta_\parallel (0)=0.1$, three values of initial $\kappa (0)$ = 2.6 (left), 4.0 (middle), and 8.0 (right), and for six different (initial) anisotropies, $T_\perp/T_\parallel (0)$, shown in various shades of brown. 
The first line (top) indicates the variation in $\kappa$, and the following ones indicate the variations in key parameters $\beta_\parallel (t)$ and $A (t)=T_\perp (t)/T_\parallel (t)$, and variations in wave properties $\delta B^2 (t)/B_0^2$ and $\gamma_{\rm max} (t)$, for both cases in which $\kappa$ varies in time and ones in which $\kappa$ is constant.}
\end{figure*}

\begin{figure*}[t!] \centering
\includegraphics[width=0.9\textwidth]{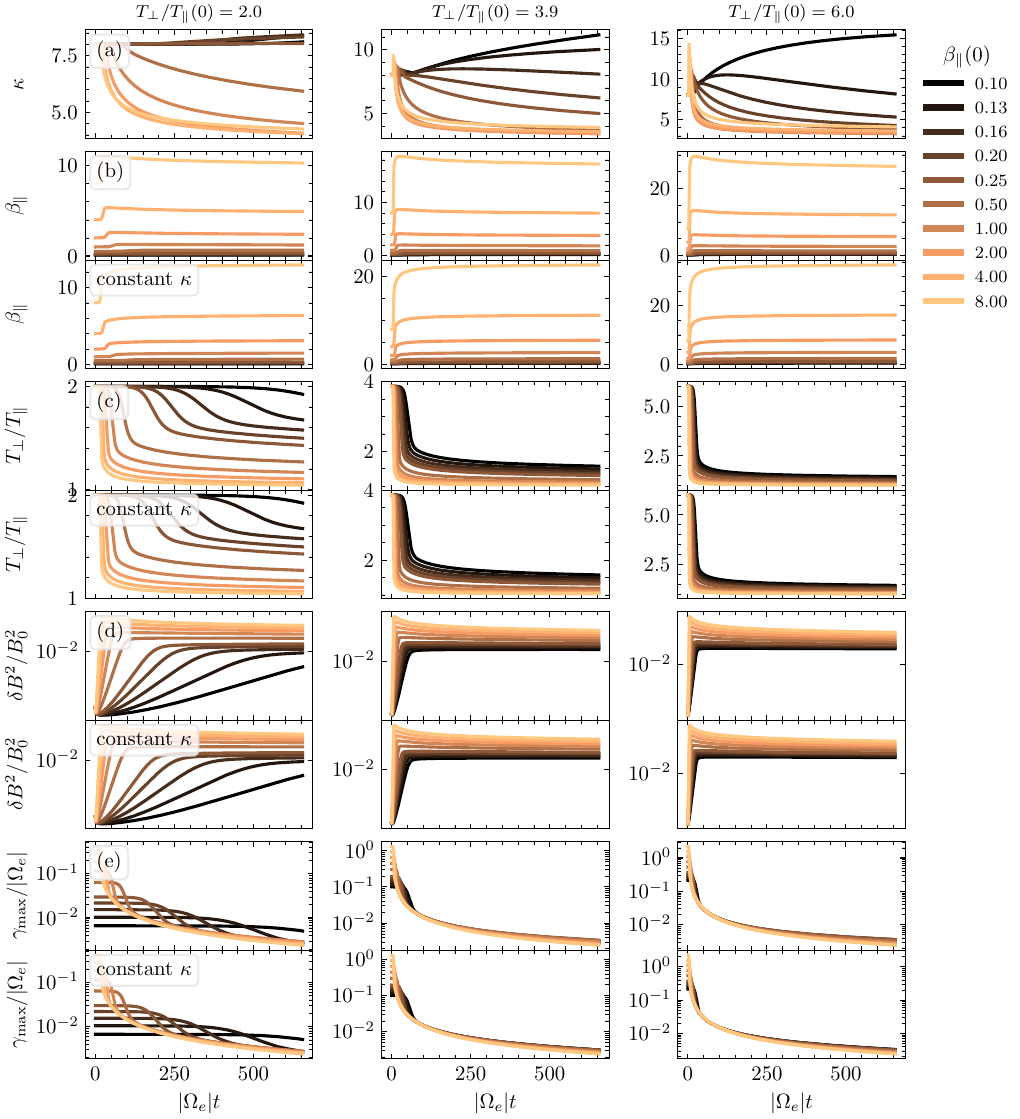}
\caption{\label{fig:3} QL runs with the same initial $\kappa (0)=8.0$, three values of initial anisotropy $A (0)$ = 2.0 (left), 3.9 (middle), and 6.0 (right), and for ten different (initial) values of $\beta_\parallel (0)$ shown in various shades of brown. 
The first line (top) indicates the variation in $\kappa$, and the following ones indicate the variations in key parameters $\beta_\parallel (t)$ and $A (t)=T_\perp /T_\parallel (t)$, and variations in wave properties $\delta B^2 (t)/B_0^2$ and $\gamma_{\rm max} (t)$, for both cases in which $\kappa$ varies in time and ones in which $\kappa$ is constant.}
\end{figure*}

Figures~\ref{fig:1}--\ref{fig:3} present the QL solutions in detail, such as the time variations in the properties of the unstable EMEC waves, such as the maximum growth rate, $\gamma_{\rm max}/ |\Omega_e|$ (normalized by the electron gyrofrequency), and the wave magnetic energy density, $\delta B^2/B_0^2$ (normalized by the regular magnetic energy density), as well as the variations in the key electron parameters on which these instabilities depend, such as the (parallel) beta parameter, $\beta_\parallel (t)$, the temperature anisotropy, $A (t)=T_\perp/T_\parallel (t)$, and of course the parameter $\kappa (t)$. The new results obtained with a more realistic QL model, in which the $\kappa$ parameter is allowed to vary self-consistently over time, are presented and compared with those from the simplified QL model, in which $\kappa$ is held constant.

\begin{figure*}[t!] \centering
\includegraphics[width=0.9\textwidth]{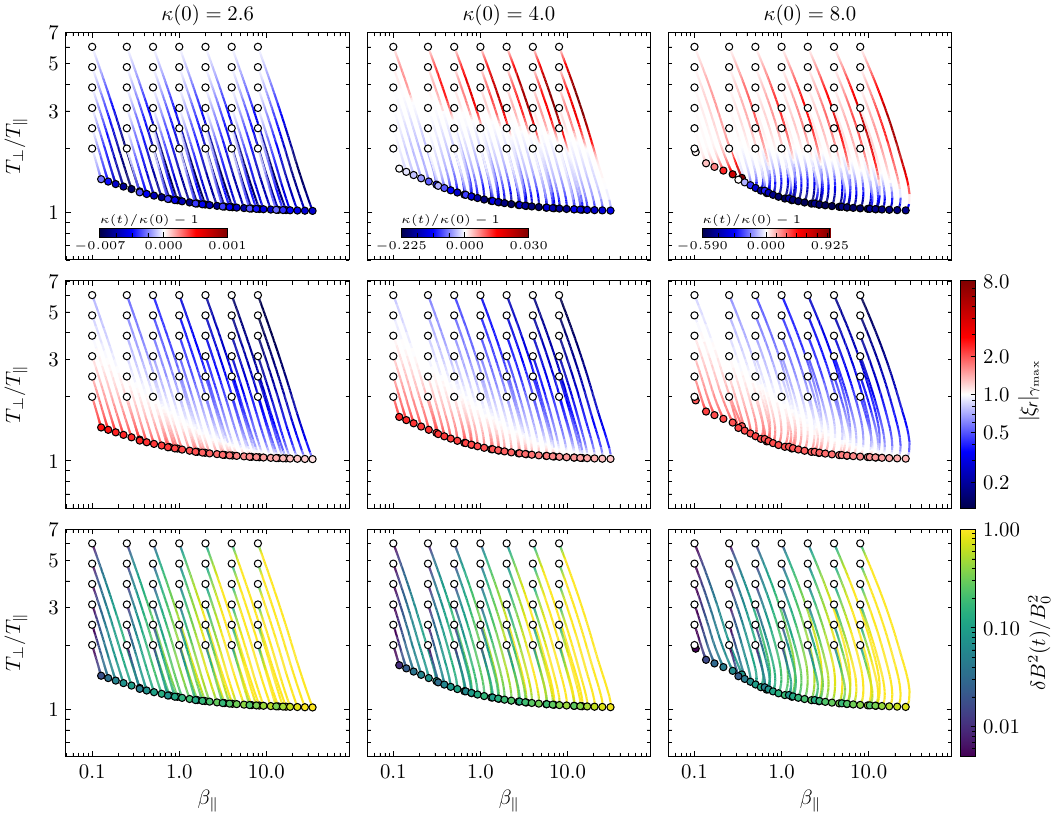}
\caption{\label{fig:4} QL dynamic paths in diagrams of $T_\perp/T_\parallel$ vs. $\beta_\parallel$, showing for each run time variations (color coded) in $\kappa(t)$ (top), the resonance factor (middle, see text), and the wave energy density (bottom). Each column groups runs with the same initial $\kappa (0)$ = 2.6 (left), 4.0 (middle), and 8.0 (right).}
\end{figure*}

Figure~\ref{fig:1} displays QL runs with the same initial anisotropy $A(0)=2$, and each column corresponds to three different values of $\beta_\parallel (0) =$ 0.10 (left), 0.50 (middle), and 8.0 (right).  The solutions represented in eight shades of brown correspond, from the darkest to the lightest, to different values of initial $\kappa (0) = 2.6, 2.8, 3.0, 4.0, 5.0, 6.0, 7.0$, and 8.0.
The difference from simplified QL models is the variation in time of the parameter $\kappa$, which may eventually decrease if $\beta_\parallel (0)$ is not very small (second and third columns), and also if the initial value of $\kappa (0)$ is large enough. The larger the $\beta_\parallel (0)$, the lower the initial value of $\kappa(0)$ affected by the decrease in time.
The QL decrease in $\kappa(t)$ over time does not indicate a relaxation toward thermodynamic equilibrium, characterized by quasi-Maxwellian distributions with $\kappa$ (much) larger than the initial one, but an increase in high-energy suprathermal tails. Whistler waves generated by the instability can contribute to electron energization, enhancing suprathermal tails \citep{Ma-Summers-1999, Vocks-etal-2008}, a sufficiently robust effect captured by simulations \citep{Lazar-etal-2017b, Lazar-etal-2018} and also indicated by the first QL models with zero-order approaches for the temporal variation in the kappa parameter \citep{Moya-etal-2021}.

Figure~\ref{fig:2} shows QL runs with the same initial $\beta_\parallel (0)=0.1$ and each column corresponds to three different initial values of $\kappa (0) = 2.6$ (left), 4.0 (middle), and 8.0 (right). The solutions in eight shades of brown correspond, from the darkest to the lightest, to different values of initial anisotropy $ A(0) = T_\perp / T_\parallel (0) = 2.0, 2.5, 3.1, 3.9, 4.8$, and 6.0. In this case, the temporal evolutions obtained for the properties of the wave fluctuations and for the key parameters describing the anisotropic electrons show no significant differences between the models with $\kappa =$~constant and those with variable $\kappa$. However, it should be noted that when $\kappa$ can vary, it does so in both directions. When the initial value of $\kappa(0)$ is not very high, as in the first two cases, $\kappa (0)=2.6$ and 4, we obtain a decrease in $\kappa$; that is, an increased suprathermalization, as was also found in the previous cases. However, it should be noted that these variations in $\kappa(t)$ are much smaller than in Fig.~\ref{fig:1}. In contrast, in the latter case, when the initial suprathermalization is reduced, with $\kappa(0) = 8$, this parameter tends to increase further, as evidence of the relaxation of the initial distribution through thermalization. This evolution is apparently more robust in the case of small initial anisotropies, $T_\perp/T_\parallel (0) < 4$, while for larger anisotropies the initial increase in $\kappa$ saturates to a maximum (around $\kappa = 10$), after which it starts to decrease, most likely under the action of an increased level of fluctuations generated by instability in this case.

In Fig.~\ref{fig:3} we plot QL runs with the same initial $\kappa (0)=8$ and each column corresponds to three different initial anisotropies: $A (0) = 2.0$ (left), 3.9 (middle), and 6.0 (right). The solutions in eight shades of brown correspond, from the darkest to the lightest, to different values of initial $ \beta_\parallel(0) = 0.1, 0.13, 0.16, 0.20, 0.25, 0.50, 1.00, 2.00, 4.0$, and 8.0.
As in Fig.~2, the variations in $\kappa(t)$ are similarly non-monotonic, but show different temporal profiles very sensitive to the initial values of the parameters. In this case, the initial $\kappa(0) = 8$ is already large, but $\kappa(t)$ can still increase over time, even for larger anisotropies, if the initial $\beta_\parallel (0)$ is low enough. If the initial anisotropy is not very low, for example $A(t) = 3.9$ and 6.0, the increase in $\kappa (t)$ occurs in longer runs, after a short but significant drop. In general, these increases are not very steep, and the final values reached at saturation do not necessarily exceed the initial ones by much.

\begin{figure}[t!] \centering
\includegraphics[width=0.9\linewidth]{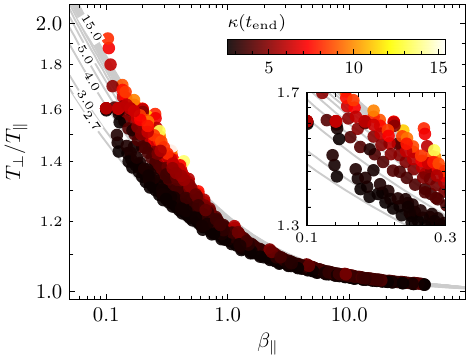}
\caption{\label{fig:5} Diagram of $T_\perp/T_\parallel$ vs. $\beta_\parallel$, showing final values of $\kappa(t_{end})$ (color-coded) obtained for the final states (filled circles), which reach the instability thresholds (light gray, $\gamma_{\rm max}/|\Omega_e|= 5 \times 10^{-3}$) predicted by linear theory for different initial values of $\kappa(0)$ (as is partially indicated).}
\end{figure}

\begin{figure*}[t!] 
  \centering
    \includegraphics[width=0.9\textwidth]{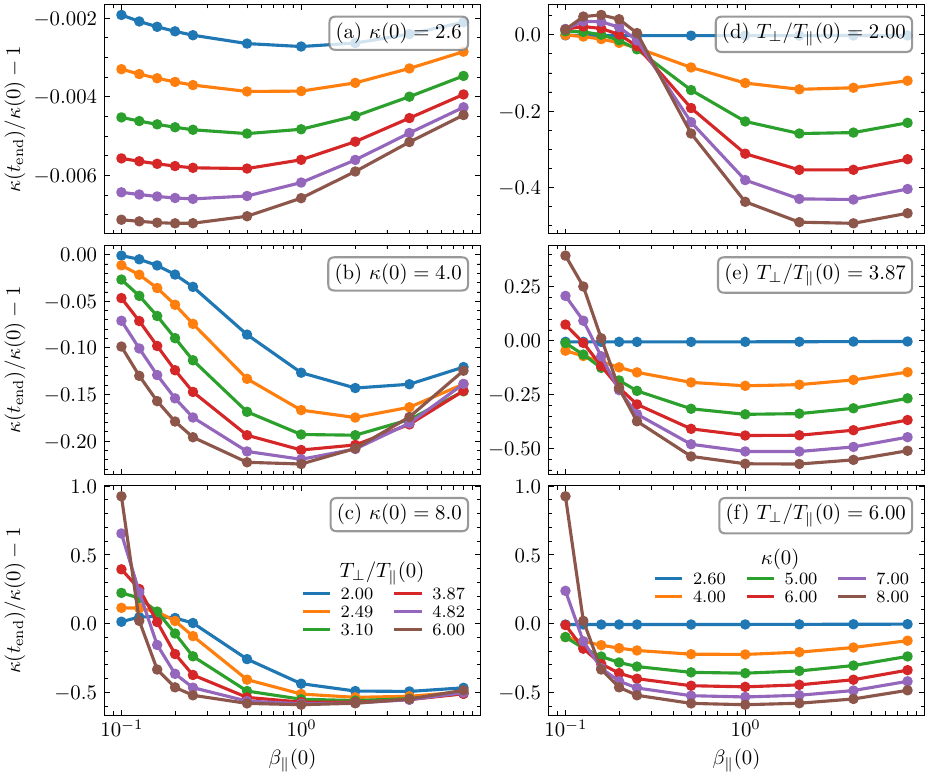}
    \caption{\label{fig:6} Relative variation in final $\kappa (t_{end})$ with respect to the initial values, as a function of initial values for $\kappa$ (panels a to c), $T_\perp/T_\parallel (0)$ (panels d to f), and $\beta_\parallel (0)$ (all panels).}
\end{figure*}

To these details, we add summarizing graphs in Figs.~\ref{fig:4}, \ref{fig:5}, and \ref{fig:6} that provide insights from the long runs, including those specific to instability saturation at the end of the QL simulations.
Thus, Fig.~\ref{fig:4} displays diagrams with the dynamic paths of the complete QL evolutions as a function of the main kinetic parameters, $T_\perp/T_\parallel$ and $\beta_\parallel$. The quasi-stationary states reached at saturation correspond to an instability threshold of $\gamma_{\rm max} / |\Omega_e| = 5 \times 10^{-3}$. For the same dynamic paths, we present with different color codes the relative variation in the parameter $\kappa(t)$ (top), the variation in $|\xi_r = \omega_r - |\Omega_e|/(k \theta_\parallel)|$, and the temporal variation in the magnetic wave energy density $\delta B^2 (t) / B_0^2$.
The top panels provide a comprehensive picture of the regimes in which the parameter $\kappa (t)$ decreases or increases. Its increase, meaning the relaxation of the electron distribution toward less suprathermal states, is limited only to regimes with sufficiently small $\beta_\parallel < 0.3$, and only when the initial value of $\kappa$ is sufficiently large, such as in the third case with $\kappa (0) = 8$. In these regimes, the levels of wave fluctuations induced by instability remain low, as is shown in the corresponding bottom panel.
Otherwise, for $\beta_\parallel > 0.3$, the variation in $\kappa$ is non-monotonic with an initial increase followed by decreases. $\xi_r$, as the real part of the argument of the plasma dispersion function, is an indicator of the resonant wave-electron interaction, which becomes maximum when $|\xi_r| \sim 1$ (see the narrow white bands), involving particles with a velocity close to the thermal velocity. $|\xi_r| < 1$ (bluish) implies the resonance of the core electrons that damp rather than amplify the waves, while $|\xi_r| > 1$ (reddish) signifies the resonance of electrons from the high-energy (suprathermal) tails that amplify the waves \citep{Lazar-etal-2019, Lazar-etal-2022}. At small anisotropies near the threshold, the suprathermal populations are highly resonant. This is also the case for regimes with low $\beta_\parallel < 1$, where an increase in the parameter $\kappa$ may eventually occur.
In addition, the bottom panels show that, as was expected, an increased wave energy density is obtained for large values of the anisotropy and beta parameters.

Figure~\ref{fig:5} uses the same diagram of ($T_\perp/T_\parallel$ vs. $\beta_\parallel$) states to show only the final states, with filled circles of different colors depending on the final value of the parameter $\kappa (t_{end})$. These are aligned with the corresponding EMEC thresholds obtained for different initial values of $\kappa(0)$, some of which are indicated to mark their increasing order. If large final values of $\kappa (t_{end})$ are obtained, it is clear that these are specific to regimes with low $\beta < 1$, but also to sufficiently high initial values of $\kappa(0)>5$. Otherwise, we are dealing with an increased suprathermalization over time; that is, a decrease in $\kappa$ to values of $\kappa(t_{end}) < 5$, also associated with the lowest anisotropies in the quasi-stationary final states. These results are in complete agreement with previous analyses, both in linear and QL theory and in numerical simulations \citep{Lazar-etal-2019, Lazar-etal-2022, Moya-etal-2021}.

Figure~\ref{fig:6} outlines the total relative variations obtained for the parameter $\kappa$ after saturation of the EMEC instability. The six panels present these relative variations as a function of the initial conditions characterized by $T_\perp/T_\parallel (0)$, $\beta_\parallel (0)$, and $\kappa(0)$. Positive relative variations are found only in panels (c), (e), and (f), which are specific to the low regimes $\beta_\parallel (0) < 0.3$. In all other cases, the final value of the kappa parameter is less than the initial value, $\kappa (t_{end}) < \kappa(0)$, and this behavior strongly depends on the initial conditions ($T_\perp/T_\parallel (0)$, $\beta_\parallel (0)$), which is subsequently related to the amount of free energy available to generate electromagnetic waves. Thus, it should be noted that, under the effect of the EMEC instability, which increases wave fluctuations, the predominance of regimes in which the parameter $ \ kappa$ decreases also enhances the suprathermalization of electrons. 

\begin{figure}[t!] 
  \centering
 \includegraphics[width=0.9\linewidth]{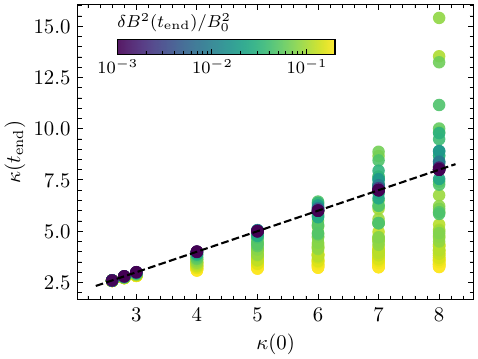}
    \caption{\label{fig:7} Scatter plot showing the final value of the kappa parameter, $\kappa (t_{end})$, as a function of the initial value. Each dot corresponds to a given QL run. The color bar represents the level of magnetic energy at the end of each calculation. The dashed line corresponds to the identity.}
\end{figure}

To further explore the connection between the decrease in kappa and the increase in the level of wave fluctuations in the system, Fig.~\ref{fig:7} shows a scatter plot of $\kappa (t_{end})$ as a function of $\kappa(0)$. In the figure, each dot corresponds to a given QL run, the color represents the level of magnetic energy at the end of each calculation $t=t_{end}$, and the dashed line corresponds to the identity. From the figure, we can see a clear correlation between the level of magnetic fluctuations and the evolution of the kappa index, quantifying the trends shown in Fig.~\ref{fig:4}. The larger the level of magnetic energy at the end of the QL run, the more pronounced the departure of the kappa parameter from its initial value. These results indicate that the relaxation of instabilities through the generation of magnetic fluctuations almost always affects the kappa value. For a large initial kappa value ($\kappa(0) >6$), the generation of electromagnetic waves induces an increase in $\kappa$, suggesting that the relaxation of the EMEC instability induces a thermalization of the VD.

However, for $\kappa < 6$, the relaxation of the EMEC instability usually results in kappa decreasing, meaning that the VD moves away from the Maxwellian limit, which defies the intuition that equilibrium corresponds to Maxwellian distributions. These small kappa regimes are ubiquitous to the solar wind~\citep{Lazar-etal-2017, Lazar-etal-2020, Eyelade-etal-2025} and planetary environments, such as the Earth's magnetosphere~\citep{Espinoza-etal-2018, Eyelade-etal-2021}, where the beta plasma parameter for electrons is generally sufficiently large, $\beta_e > 0.3$ \citep{Stverak-etal-2008, Wilson-etal-2019a, Wilson-etal-2019b, Abraham-etal-2022}. In situ observations, particularly in the solar wind, reveal mainly quasi-stationary states in the vicinity of the lowered EMEC thresholds, whose suprathermalization increases with heliocentric distance as the parameter $\kappa$ decreases toward 1~AU and beyond.


%
\section{Conclusions}
\label{sec:conclusions}

In this paper we have proposed a new dynamical model for the QL analysis of EMEC (or whistler) instabilities triggered by anisotropic electrons with bi-Kappa VDs, as observed in space plasmas. In contrast to the first QL models, which constrain the parameter $\kappa$ (or the distribution's functional form) to remain constant, the new approach allows $\kappa$ to vary self-consistently over time. \cite{Moya-etal-2021} proposed a first zero-order model with variable $\kappa$ in which they nevertheless constrained the low-energy core populations (described by the Maxwellian limit $\kappa \to \infty$ of Kappa distributions) to remain isolated from the suprathermal ones (in the high-energy tails of Kappa distributions). In the new QL approach, we have also overcome this constraint by completing the QL equations with those for the fourth-order moment (kurtosis).

The results refine and extend the previous ones, this time making a clear distinction between alternative regimes that lead to either a decrease or an increase in the parameter $\kappa$ during the stabilization of the instability. In agreement with the zero-order approach, a decrease in $\kappa$, i.e., an excess of suprathermalization, prevails. However, this time, the (initial) regimes in which the VDs can relax to a quasi-Maxwellian form due to the increase in $\kappa$ are also identified. These (initial) conditions are generally restricted to low beta parameters, but not to very small initial values for $\kappa \gtrsim 5$.
The main conclusion is that, in general, the bi-Kappa electron VDs undergo only a partial relaxation under the action of the EMEC instability. Only the temperature anisotropy decreases, while wave fluctuations result in excess and maintain but also enhance the suprathermal component, as is evidenced by the decrease in $\kappa$. In other words, for small $\kappa$ regimes, during the relaxation of the EMEC instability, electromagnetic instabilities are produced as temperature anisotropy decreases due mainly to resonant interactions with the quasi-thermal core (see the central panel of Fig.~\ref{fig:4}). Subsequently, the self-generated waves can interact with suprathermal particles, generating energetic power-law tails represented by a small $\kappa$ value. Additionally, the relations obtained between magnetic fluctuations and the kappa index suggest that a fixed value of $\kappa$ is not an adequate strategy for studying kinetic instabilities in nonthermal plasmas. Thus, the consideration of a dynamic kappa should be customary for the study of nearly collisionless space plasmas.

That said, it is worth mentioning that the whistler mode can be driven by temperature anisotropies and/or heat flux. In the case of the heat-flux-driven instability, the free-energy source corresponds to the asymmetry of the electron velocity distribution function (VDF), quantified by the third moment of the VDF, as is shown in \cite{Tong2019}. In the solar wind, this is usually provided by inherent asymmetry~\citep{,Zenteno2021model,zenteno2022role,Zenteno2023interplay} or the relative drift between the core and halo \citep{Vasko2020} or strahl \citep{Lopez2020b}. In this case, to focus on the temperature-anisotropy-driven EMEC instability, our model assumes a single, symmetric Kappa distribution, such that all odd moments of the VDF are identically zero. It would be interesting to address the role of asymmetry and heat flux for the evolution of Kappa distributions during the relaxation of the whistler heat-flux instability. In such a case, a model based on an asymmetric VDF will allow for the third-order moment to be considered in order to complement the kurtosis equation, or replace it, under the assumption of a constant electron drift speed along the magnetic-field direction. However, analyzing such a model is beyond the scope of the current manuscript and will be left for future work.

In summary, our results suggest that the microscopic and macroscopic characteristics of the plasma, and the fine structure of electromagnetic field fluctuations, waves, and turbulence, strongly depend on the shape of the VD and its evolution, and vice versa, particularly after excitation and during relaxation processes associated with collisionless dissipation. Our results already show preliminary agreement with in situ observations in the solar wind and the magnetosphere, suggesting that the new QL model could provide a theoretical basis for explaining similar kinetic instabilities in natural plasmas with bi-Kappa-type distributions. This hypothesis may benefit in the future from confirmations through numerical simulations and a detailed analysis of anisotropic distributions during relaxation under the action of instabilities.

\begin{acknowledgements}
The authors acknowledge support from the University of Chile, Ruhr-University Bochum, and the Katholieke Universiteit Leuven. Portions of this research were supported by the International Space Science Institute (ISSI) in Bern, through the ISSI International Team project 24-612: Excitation and Dissipation of Kinetic-Scale Fluctuations in Space Plasmas. These results were also obtained in the framework of the projects C16/24/010 (C1 project Internal Funds KU Leuven), G002523N (FWO-Vlaanderen), 4000145223 SIDC Data Exploitation (SIDEX2), ESA Prodex, and Fondecyt ANID Chile No. 1240281 (PSM) and No. 1251712 (RAL). SP is funded by the European Union (ERC, Open SESAME, 101141362). Views and opinions expressed are, however, those of the author(s) only and do not necessarily reflect those of the European Union or the European Research Council. Neither the European Union nor the granting authority can be held responsible for them.

\end{acknowledgements}

\bibliographystyle{aa} 
\bibliography{biblio}

\end{document}